\documentclass[preprint
]{elsarticle}

\usepackage[]{graphicx}
\usepackage{epstopdf}
\usepackage{amsmath,amssymb}
\usepackage{upgreek}
\usepackage[utf8]{inputenc}
\usepackage[colorlinks, linkcolor=blue, citecolor=blue, urlcolor=blue,breaklinks=true]{hyperref}
\usepackage{soul,physics,enumitem,cleveref,xspace}

\providecommand{\HnH}{\ensuremath{H_\text{nH}}}
\providecommand{\nTh}{\ensuremath{n_\text{Th}}}
\providecommand{\Dp}{\ensuremath{\Delta p}\xspace}
\providecommand{\Dpc}{\ensuremath{\Dp_\text{c}}\xspace}

\providecommand{\expprob}[4][\big]{\mathbb #2\qty #1[#3\, #1|\,#4]}

\graphicspath{{./figures/}}

\begin{document}

\title{The Monte Carlo wave-function method: a robust adaptive algorithm and a study in convergence}
\author[wrc,elu]{M. Kornyik}
\ead{kornyik.miklos@wigner.mta.hu}
\author[wrc]{A. Vukics\corref{cor1}}
\ead{vukics.andras@wigner.mta.hu}

\cortext[cor1]{Corresponding author}

\address[wrc]{Wigner Research Centre for Physics, H-1525 Budapest, P.O. Box 49., Hungary}
\address[elu]{Eötvös Loránd University, H-1117 Budapest, Pázmány Péter stny. 1/C., Hungary}

\date{\today}

\begin{keyword}
Monte Carlo wave function\sep quantum jumps\sep stochastic simulations\sep open quantum systems\sep Markov approximation\sep adaptive stepsize

\PACS 02.70.Ss\sep 33.80.-b\sep 42.50.Lc
\end{keyword}

\begin{abstract}
We present a stepwise adaptive-timestep version of the Quantum Jump (Monte Carlo wave-function) algorithm. Our method has proved to remain robust even for problems where the integrating implementation of the Quantum Jump method is numerically problematic. The only specific parameter of our algorithm is the single \emph{a priory} parameter of the Quantum Jump method, the maximal allowed total jump probability per timestep. We study the convergence of ensembles of trajectories to the solution of the full master equation as a function of this parameter. This study is expected to pertain to any possible implementation of the Quantum Jump method.
\end{abstract}

\maketitle

\tableofcontents

\section{Introduction}

The Quantum Jump (Monte Carlo wave-function – MCWF\footnote{The use of the term “wave function” is not completely correct in this version of the name because the method is generally applicable for state vectors and not only for those expanded in space (which are the wave functions). Nevertheless, we continue using this customary acronym.}) method has been around since at least the late 1980s, the notion of quantum jumps being introduced in connection with intermittent fluorescence \cite{Plenio1998} in works like \cite{Diosi1985,Javanainen1986}. The first versions of implementable algorithms were published in the early 1990s \cite{Dum1992,Dalibard1992}. In a parallel development, another kind of quantum trajectory methods, the quantum state diffusion has been worked out \cite{Gisin1992}.

The Quantum Jump method can be put forward with two distinct motivations:
\begin{description}
 \item[As a computational tool] to unravel the quantum master equation into a set of quantum trajectories in order to reduce the dimensionality of the numerical problem to make larger systems tractable. In this case, it is not necessary to endow the individual trajectories with any physical meaning. 
 \item[As a physical model] to reflect the behavior of single realizations of small quantum systems. While quantum mechanics was originally conceived to describe ensembles, with single ions in Paul traps \cite{Cook1985} being the first examples it has in the last few decades become possible to study single realizations. In this case, the individual trajectories can be considered physical, and they will depend on the way the system is observed, in accordance with the lore of quantum measurement.
\end{description}
While the benefit in terms of net computational resources as compared to direct master equation simulation is not clear cut \cite{Breuer1997}, since too many trajectories may be needed for acceptable statistics; in realistic situations the system is often so big that even a single copy of the full density matrix exceeds memory limits. Then, in the ergodic case, it is still a possible solution to content oneself with finding the steady state via time averaging along a single long trajectory. Already a single electromagnetic mode coupled to a few-level system (like in cavity quantum electrodynamics) can easily fall into this category under realistic conditions \cite{Dombi2013,Dombi2015}, but in this way it was possible to study a system consisting of two atoms coupled to a single mode \cite{Vukics2007}, or two-modes (ring cavity) coupled to a single atom \cite{Niedenzu2010}. Ergodicity could be utilized even in the case of a system featuring two distinct semiclassical attractors \cite{Fink2017}, although with heavy computational cost. Recently, quantum trajectories have been applied also in quantum many body context \cite{Daley2014,Kirton2017}, sometimes together with tensor-network techniques.

Adaptive algorithms are very important in dynamical simulations as in general there is no way to predict an optimal stepsize, which even varies along the trajectory evolution. While in the case of deterministic problems (ordinary differential equations – ODEs), seasoned robust generic algorithms exist, the same is not true for stochastic problems (stochastic differential equations – SDEs). Such dynamics and their numerical simulation have been intensively researched during the last decades, numerous excellent papers and books can be found in the literature giving conditions on strong and weak convergence, stability, and also rates of convergence of the discretized solution \cite{buckwar2011runge,buckwar2011numerical,clark1980maximum,talay1994numerical}. The simplest such method is the Euler-Maruyama scheme which is basically the extension of the  explicit Euler method well-known from the theory of ODEs (for a more detailed description of the various numerical methods the reader is referred to \cite{kloeden2012numerical}). As opposed to the deterministic case, where the order of (global) convergence is 1, the order of strong convergence in the case of the Euler-Maruyama method is only 1/2. Generally, no numerical method based only on an approximation of the Brownian motion can guarantee an asymptotic convergence rate higher than that \cite{gaines1997variable}. For higher orders of strong convergence (\(n/2\) with \(n\geq2\)), the Itô-Taylor expansion yields an answer, which involves approximation of Lévy areas, i.e. integrals of Brownian motion. Unfortunately, due to the properties of the Itô integral, these schemes are more complicated than their deterministic counterparts. Usually, higher order methods are computationally very expensive, and in order to save computational time, variable stepsize for lower order methods was introduced and various results on convergence rate and the optimal choice of the stepsize were published as well \cite{ilie2015adaptive}.

These developments are not directly relevant to MCWF because it does not consist of the integration of an SDE, but they may be utilized for Quantum State Diffusion, which does have the form of an SDE – to our knowledge, higher-order methods have not yet been tried in this case. The MCWF can be described as an SDE that consists of an ODE driven by a general Poisson-process.

In this paper, we present a stepwise adaptive algorithm to simulate this process that is by principle more robust than the popular implementation of the MCWF method \cite{Breuer1995,Homa2017} that we denote the “integrating method” in this paper, and that is used e.g. in the popular QuTiP package \cite{Johansson2012,Johansson2013}. The increased robustness, whose main reason is that the stepwise algorithm does not depend on an algorithm for retrieving a past jump time instant, comes at the cost of some reduction of efficiency, which however we will argue to be marginal in most usecases we encountered so far. The algorithm here presented has been used in C++QED \cite{VukicsCppQEDa,VukicsCppQEDb,Sandner2014CppQED} since the inception of that framework, both the algorithm and framework having been originally developed for the demanding problem of simulating motional quantum degrees of freedom expanded in momentum space \cite{Vukics2005a,Vukics2005b,Vukics2007,Vukics2009,Schulze2010,Niedenzu2010,Niedenzu2012,Sandner2013,Winterauer2015}. We will argue that in this field, the robustness of our algorithm over the integrating one is especially expressed.

Apart from the parameters governing the precision of the ODE integration, the stochastic part of our algorithm is governed by the single additional dimensionless parameter \(\Delta p\): the maximal allowed total jump probability per step. Another aspect of the present paper is a study of convergence of ensembles of MCWF trajectories to the solution of the density operator as a function of this parameter. This being also the most important \emph{a priori} parameter of the MCWF method, this convergence study is not specific to our algorithm, but pertains to any possible implementation of the MCWF method, including the integrating one.

The paper is structured as follows: \Cref{sec:primordialMCWF} describes the original MCWF algorithm and its issues that we set out to address with our adaptive algorithm. The latter is described in detail in \Cref{sec:updatedMCWF}. \Cref{sec:convergence} is devoted to the numerical investigation of the convergence of ensembles of trajectories computed by our adaptive algorithm to the solution of the master equation. We will see (\Cref{sec:pureDpControl}) that the problem of a finite-temperature harmonic oscillator mode driven purely by photon exchange with the bath is the most demanding of the simple generic examples, due to a kind of bosonic enhancement of the noise. In the case of a nontrivial Hamiltonian (\Cref{sec:contention}), a contention between the ODE stepsize-control heuristic and our superimposed heuristic of jump-probability control takes place. In \Cref{sec:outlook}, we compare our algorithm to the integrating algorithm \cite{Breuer1995,Homa2017} of MCWF evolution, showing some realistic usecases that favor our method (\Cref{sec:favoringUsecases}). Finally, we share some insights about sampling and time averaging (\Cref{sec:sampling}).

\section{Primordial MCWF and its critique}
\label{sec:primordialMCWF}
The MCWF method aims at unravelling a master equation into a statistical ensemble of stochastic quantum trajectories, whose initial condition is a corresponding ensemble of state vectors that appropriately samples the initial density operator. Besides being a useful theoretical tool for reducing the dimensionality of the numerical problem, it furthermore reflects the – physically unrealistic – situation when an experimenter is in full control of any single copy (realization, experimental run) of the physical system, both in terms of controlling possible pure-state initial conditions and observing all the possible quantum jumps (e.g. photon decays) the system undergoes over time.

A master equation in the Born-Markov approximation in the most general – so called Lindblad – form reads:
\begin{subequations}
\begin{equation}
\label{eq:master}
\dot\rho=\frac 1{i\hbar}\commutator{H}{\rho}+\sum_m\qty(J_m\rho J_m^\dagger-\frac12\acomm{J_m^\dagger J_m}{\rho})\equiv2\,\mathfrak{Re}\qty{\frac\HnH {i\hbar}\rho}+\sum_m J_m\rho J_m^\dagger,
\end{equation}
where with the second equality we have defined the non-Hermitian “Hamiltonian”
\begin{equation}
\label{eq:nHH}
\HnH\equiv H-\frac{i\hbar}2\sum_m J_m^\dagger J_m.
\end{equation}
\end{subequations}
The \(J_m\) operators are called \emph{quantum jump} – or \emph{Lindblad,} or \emph{reset,} or \emph{collapse} – operators, and their maximum number is one less than the square of the dimension of the physical system.

The state-vector initial condition \(\ket{\Psi(0)}\) of a single trajectory is taken from an ensemble that appropriately samples the initial density operator \(\rho(0)\) (in general, we need many state vectors from this ensemble and many trajectories \emph{for each state-vector initial condition}). In its original form, the MCWF algorithm to evolve \(\ket{\Psi(t)}\) to \(\ket{\Psi(t+\delta t)}\) can be listed as follows.
\begin{enumerate}
 \item The state vector is evolved according to the nonunitary dynamics
 \begin{subequations}
 \begin{equation}
 i\hbar\frac{d\ket{\Psi}}{dt}=\HnH\ket{\Psi}. 
 \end{equation}
 In the next derivations we will neglect the terms including $\delta t$ of order higher than 2. Then
 \begin{equation}
 \ket{\Psi_{\text{nH}}(t+\delta t)}=\qty(1-\frac{i\HnH\,\delta t}\hbar) \ket{\Psi(t)}. 
 \end{equation}
 Since \HnH\ is non-Hermitian, this new state vector is not normalised. The square of its norm reads
 \begin{equation}
 \bra{\Psi_{\text{nH}}(t+\delta t)}\ket{\Psi_{\text{nH}}(t+\delta t)}=\bra{\Psi(t)}\qty(1+\frac{iH^\dag_{\text{nH}}\,\delta t}\hbar)\qty(1-\frac{i\HnH\,\delta t}\hbar)\ket{\Psi(t)}\equiv 1-\delta p,
 \end{equation}
 where \(\delta p\) reads
 \begin{equation}
 \delta p=\delta t\,\frac i\hbar \bra{\Psi(t)}\HnH-H^\dag_{\text{nH}}\ket{\Psi(t)}\equiv\sum_m\delta p_m,\quad\delta p_m=\delta t\,\bra{\Psi(t)} J^\dag_m J_m\ket{\Psi(t)}\geq 0.
 \end{equation}
 \end{subequations}
 Note that the timestep \(\delta t\) should be small enough that this first-order calculation be valid. Finding the appropriate \(\delta t\) is the main theme of the present paper. In particular, we require that
 \begin{equation}
 \label{eq:dpCond}
 \delta p\ll1.
 \end{equation}
 This is important in order that the probability of \emph{two jumps} occuring in the same timestep be negligible. The primordial MCWF algorithm is first order in the sense that it cannot deal correctly with events like this. Higher order MCWF algorithms have been developed \cite{Steinbach1995}, but they require a combinatorically increasing number of jump operators in order to account correctly for every possible multi-jump event.
 \item A possible quantum jump with total probability \(\delta p\). For the physical interpretation of such a jump, cf. \cite{Dum1992,Dalibard1992}. Choose a random number \(\mathfrak r\) between 0 and 1, and if \(\delta p<\mathfrak r\) – which should mostly be the case – no jump occurs and for the new normalised state vector at \(t+\delta t\) take
 \begin{equation}
 \eval{\ket{\Psi(t+\delta t)}}_\text{no jump}=\frac{\ket{\Psi_{\text{nH}}(t+\delta t)}}{\sqrt{1-\delta p}}. 
 \end{equation}
If \(\mathfrak r<\delta p\), on the other hand, a quantum jump occurs, and the new normalised state vector is chosen from among the different state vectors \(J_m\ket{\Psi(t)}\) with probability distribution \(\Pi_m=\delta p_m/\delta p\):
\begin{equation}
\label{eq:jumpEvolution}
\eval{\ket{\Psi(t+\delta t)}}_{m\text{th jump}}=\sqrt{\delta t}\frac{J_m\ket{\Psi(t)}}{\sqrt{\delta p_m}}.
\end{equation}

\end{enumerate}

These steps can be easily shown to reproduce the master-equation evolution to first order in \(\delta t\). Let us consider the delta of a state-vector diad on a single trajectory:
\begin{multline}
\label{eq:masterDerivation}
 \rho_\text{1traj}(t+\delta t)=\ket{\Psi(t+\delta t)}\bra{\Psi(t+\delta t)}\\=
 \qty(1-\delta p)\eval{\ket{\Psi(t+\delta t)}\bra{\Psi(t+\delta t)}}_\text{no jump}
 +\sum_m \delta p_m \eval{\ket{\Psi(t+\delta t)}\bra{\Psi(t+\delta t)}}_{m\text{th jump}}\\=
 \ket{\Psi_{\text{nH}}(t+\delta t)}\bra{\Psi_{\text{nH}}(t+\delta t)}+
 \delta t \sum_m J_m \ket{\Psi(t)}\bra{\Psi(t)} J_m^\dagger\\=
 \qty(1-\frac{i\HnH\,\delta t}\hbar) \ket{\Psi(t)}\bra{\Psi(t)}\qty(1+\frac{i\HnH^\dagger\,\delta t}\hbar)+\delta t \sum_m J_m\,\rho_\text{1traj}(t)\,J_m^\dagger\\=
 \rho_\text{1traj}(t)+
 \delta t\qty(\frac{\HnH}{i\hbar}\rho_\text{1traj}(t)-\rho_\text{1traj}(t)\frac{\HnH^\dagger}{i\hbar})+\delta t \sum_m J_m\,\rho_\text{1traj}(t)\,J_m^\dagger+\order{\delta t^2}\quad\blacksquare
\end{multline}

This derivation displays that the first term in the rightmost part of \cref{eq:master} describes the no-jump evolution. It is a non-Hermitian evolution, because an open system is open not only at the time instants of jumps, but always: the no-jump periods also leak information about the system. Hence, the no-jump evolution in general cannot remain Hermitian. Conversely, the second term in the same part of \cref{eq:master} alone is responsible for the quantum jumps.

\begin{figure}
 \includegraphics[width=\linewidth]{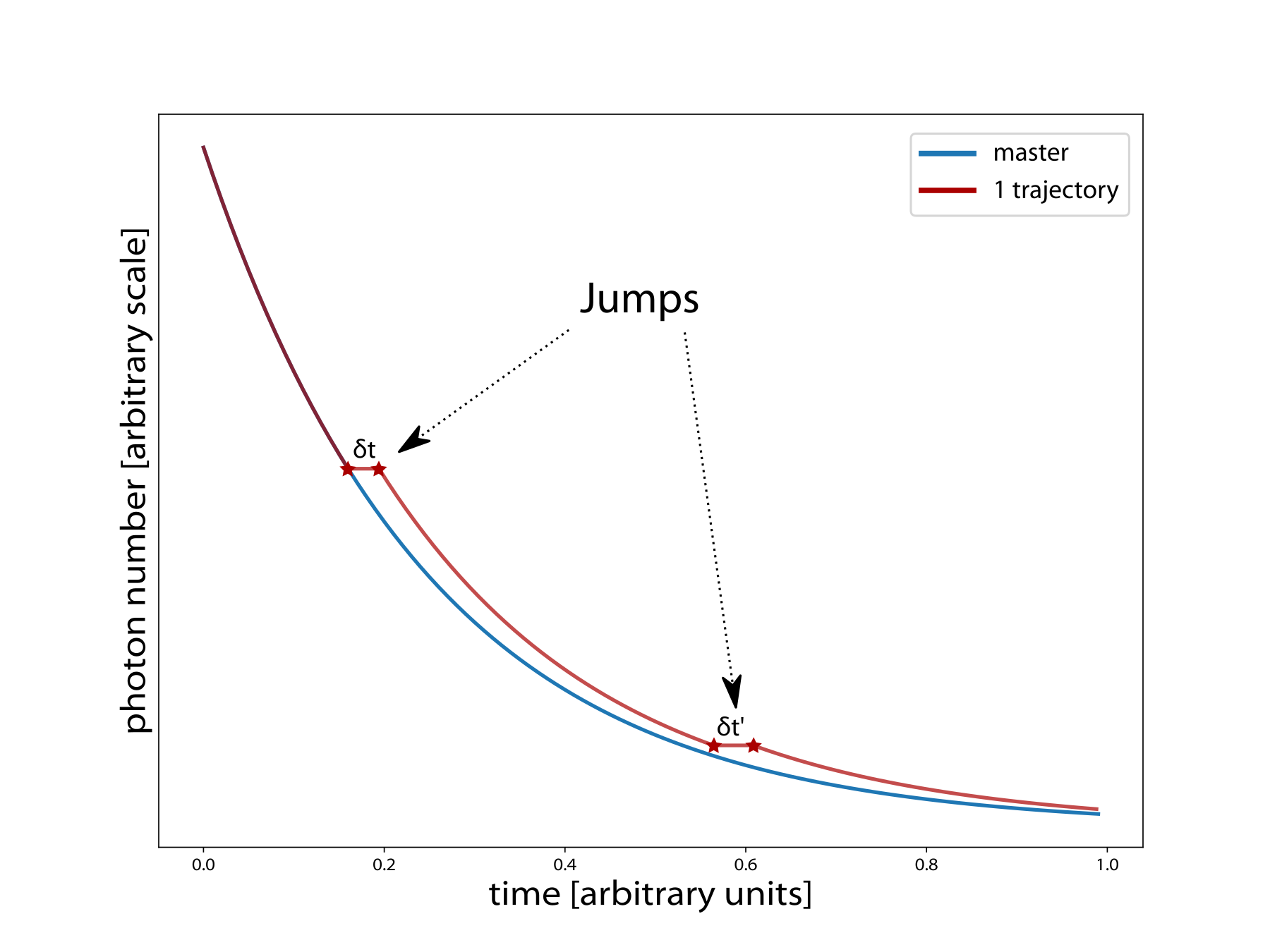}
 \caption{Cartoon of the evolution of the photon number in a single decaying harmonic-oscillator mode started from a coherent state. Since the coherent state is an eigenstate of the jump operator (cf. \Cref{sec:convergence} with \(\nTh=0\) for the mathematical scenery of this situation), the state remains coherent throughout, only with decaying amplitude. The blue line represents the correct solution obtained from the master equation. The red line is a single trajectory in the case when we allow jumps to take finite time. In this case, every jump introduces a temporal shift equal to the actual timestep with respect to the correct solution because the state is unchanged under a jump, the coherent state being eigenstate to the jump operator. Of course, in the \(\text{timestep}\to0\) limit, the correct behavior is recovered, however, a systematic error is introduced with finite stepsize by the incorrect treatment of jumps.}
 \label{fig:coherentDecay}
\end{figure}

This algorithm has several issues:
\begin{enumerate}
 \item The no-jump evolution reduces to the Euler method of ODE evolution, which is inadequate for all but the most trivial problems.
 \item The quantum jump takes finite time, since in a timestep \(\delta t\), we \emph{either} make an ODE step, \emph{or} perform a jump. This is because in the right-hand side of \cref{eq:jumpEvolution}, we use the unevolved state vector. Whether a jump should take a finite time has been discussed in the literature (cf. \cite{Plenio1998} Sec. IV.C), but here we present a strong argument that it should not.
 
 Let us consider a decaying harmonic-oscillator mode started from a coherent state. If we allowed jumps to take a finite time, then a single trajectory would deviate from the solution of the master equation as displayed in \cref{fig:coherentDecay} (cf. the figure caption for detailed explanation). Since coherent states are (the most) classical states, we do not want to allow such a deviation, since physically a classical evolution is expected for a single trajectory as well as for the master equation or any sub-ensembles.
 \item Timestep is not adaptive. This is a problem already in ODE, but in MCWF it creates the additional problem that the satisfaction of the condition \cref{eq:dpCond} remains uncontrolled.
\end{enumerate}

\section{Stepwise adaptive MCWF}
\label{sec:updatedMCWF}
While the integrating algorithm of MCWF evolution sidesteps these issues thanks to its peculiar treatment of jumps (cf. \cref{sec:outlook}), we aimed at a \emph{stepwise adaptive} algorithm that rectifies them. By ‘stepwise’ we mean that the possibility of one of the possible quantum jumps to occur is accounted for in each timestep.
\newlist{adenumerate}{enumerate}{10}
\setlist[adenumerate]{label=Ad \arabic*.}

\begin{adenumerate}
 \item Instead of the 1st order Euler ODE step, we use a higher order adaptive method \footnote{A general-purpose choice used also in the examples throughout this paper is the Runge-Kutta Cash-Karp stepper, which is fifth order with embedded fourth order error estimator.}. Regarding timestep control, such a routine expects a timestep \(\delta t_\text{try}\), which is the timestep to try in the actual step, and yields
 \begin{enumerate}
 \item \(\delta t_\text{did}\), the timestep actually performed
 \item \(\delta t_\text{next}\), which is the timestep to try in the next step
\end{enumerate}
 
When used sequentially, \(\delta t_\text{try}\) is always equated to the \(\delta t_\text{next}\) obtained in the previous step. It is important that
 \begin{equation}
  \delta t_\text{did}\leq\delta t_\text{try},
 \end{equation}
 that is, the stepper is not allowed to overshoot the suggestion obtained from the step before, while \(\delta t_\text{next}\) can be bigger than \(\delta t_\text{did}\), providing a mechanism for increasing the timestep.

\item ODE evolution is not optional during a timestep (this follows partly from item Ad 1 just above), but an ODE step is always taken, and at the end of that step, it is decided whether or not a quantum jump is taken \emph{in addition} in the same timestep. The jump itself is instantaneous.

\item Timestep is naturally adaptive stemming from item Ad 1 above, so we need to control the fulfillment of condition (\ref{eq:dpCond}). For this, we introduce a new parameter \Dp of the algorithm representing the maximum allowed total jump probability in a timestep. Clearly, we expect the algorithm to work correctly if
\begin{equation}
 \Dp\ll1.
\end{equation}
The behavior of MCWF as a function of \Dp is the main theme of the remainder of the paper.

\end{adenumerate}

\begin{figure}
 \includegraphics[width=\linewidth]{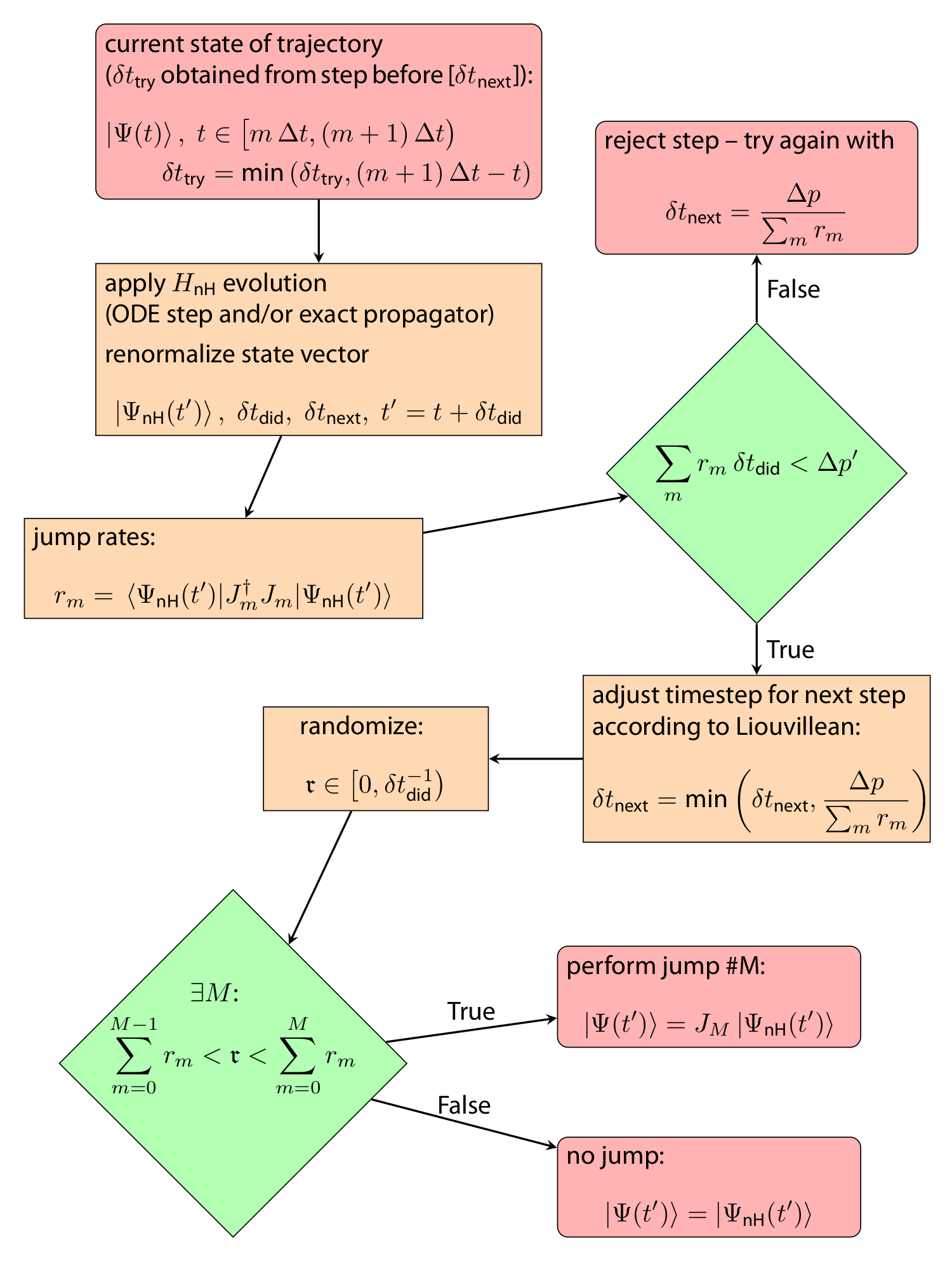}
 \caption{Flowchart describing a single step of our adaptive MCWF algorithm. Besides the physical parameters and those governing the ODE evolution, the parameters of the algorithm are \(\Delta t\), \Dp, and \(\Dp'\). The two-layer control makes sense only if \(\Dp'>\Dp\).}
 \label{fig:flowchartFull}
\end{figure}

Our adaptive algorithm can then be summed up in the form of a flowchart as in Fig. \ref{fig:flowchartFull}. Superposed on the ODE stepsize-control heuristic, we use a two-layer heuristic: upon calculating the jump rates after the \(H_\text{nH}\) evolution,
\begin{enumerate}
 \item the timestep guess fed back to the ODE stepper at the beginning of the next step is abridged by \Dp-control as:
 \begin{equation}
  \delta t_\text{next}\leq\frac{\Dp}{r_\text{tot}},
 \end{equation}
  where \(r_\text{tot}=\sum_mr_m\) is the total jump rate. The problem here is that a \Dp-overshot is handled only in the next step.
 \item if it is found that the total accumulated probability in the given step is too high, i.e. \(r_\text{tot}\,\delta t_\text{did}>\Dp'\), with some \(\Dp'>\Dp\), then the step is rejected and we go back to the beginning of the given step by restoring the state vector to a copy cached at time \(t\). In general, the internal state of the ODE stepper needs to be cached and restored as well. The next step is tried with timestep reduced as
 \begin{equation}
 \label{eq:DpControl}
  \delta t_\text{next}=\frac{\Dp}{r_\text{tot}},
 \end{equation}
\end{enumerate}

Layer 2, introduced as a safety measure, requires additional resources (although usually negligible compared to the several copies of the state vector the ODE stepper has to store internally during a step), and our experience is that its usefulness is very difficult to quantify in real-life situations. Hence, in the following we will only study the effects of the 1st layer of control, and switch off the 2nd one (this can always be done by choosing a very large \(\Dp'\) value).

One last important difference of our algorithm compared to the original must be noted: we perform an \emph{exact} renormalization of the state vector just before calculating the jump rates. Keeping the norm constant can be a very strong stabilizing condition for the ODE evolution, in certain problems that we will discuss in \cref{sec:favoringUsecases}. In these problems, experience has shown that the exact renormalization can stabilize an otherwise unstable ODE evolution, or make an otherwise very small timestep bigger. The need for exact renormalization in this sense means that the integrating method, which lets the norm evolve freely and monitors its value to determine the jumps is hindered.

It is easy to verify that the first-order-in-\(\delta t\) derivation of the master-equation evolution presented in \cref{eq:masterDerivation} is unchanged by our modifications to the algorithm, which remains of order 0.5 despite the fact that the order of the ODE stepper can be higher. This is because the handling of the quantum jumps is essentially unchanged: we allow at most one jump per timestep.

For each timestep, we define a maximum time that the stepper is allowed to reach. This is important in order to define time instants during the trajectory at which the trajectory is sure to stop. These are defined as
\begin{equation}
\text{sampling times}\equiv u\,\Delta t\qqtext{with}u\in\mathbb{N}.
\end{equation}
Hence, at these points all the trajectories of an ensemble can be brought together e.g. for ensemble averaging (due to adaptive stepping, the trajectories have different times after fixed number of steps). As we shall see below, \(\Delta t\) is an important parameter that in principle pertains to the convergence of the method.

\section{Convergence}
\label{sec:convergence}
To study the convergence properties of the adaptive MCWF as a function of \Dp, we choose a very simple system: a single harmonic-oscillator mode interacting with a finite-temperature reservoir. If \(a\) denotes the annihilation operator of the mode excitations, the jump operators read:
\begin{subequations}
\begin{align}
 J_0&=\sqrt{2\kappa\qty(\nTh+1)}\,a&\qqtext{(photon emission),}\\
 J_1&=\sqrt{2\kappa\,\nTh}\,a^\dagger&\qqtext{(photon absorption).}
\end{align} 
\end{subequations}
We set the timescale such that \(\kappa=1\).

The mode can be driven by a coherent drive, that we will consider resonant with the mode frequency, in which case we find a time-independent Hamiltonian in the frame rotating with the frequency of the drive:
\begin{equation}
\label{eq:modeHamiltonian}
 H=i\hbar\,\eta\,\qty(a^\dagger-a),
\end{equation}
where \(\eta\) is the Rabi frequency of the drive. The non-Hermitian Hamiltonian reads:
\begin{equation}
\label{eq:HnH_mode}
 H_\text{nH}=-i\hbar\,\kappa\qty(2\nTh+1)\,a^\dagger a+i\hbar\,\eta\,\qty(a^\dagger-a).
\end{equation}
The coherent drive tends to impose a coherent steady state even on a single trajectory, which is stabilized also by the photon emission (eigenstate of \(J_0\)), the photon absorption being the only mechanism acting against it.

\subsection{Pure \Dp-control}
\label{sec:pureDpControl}

\begin{figure}
 \includegraphics[width=.9\linewidth]{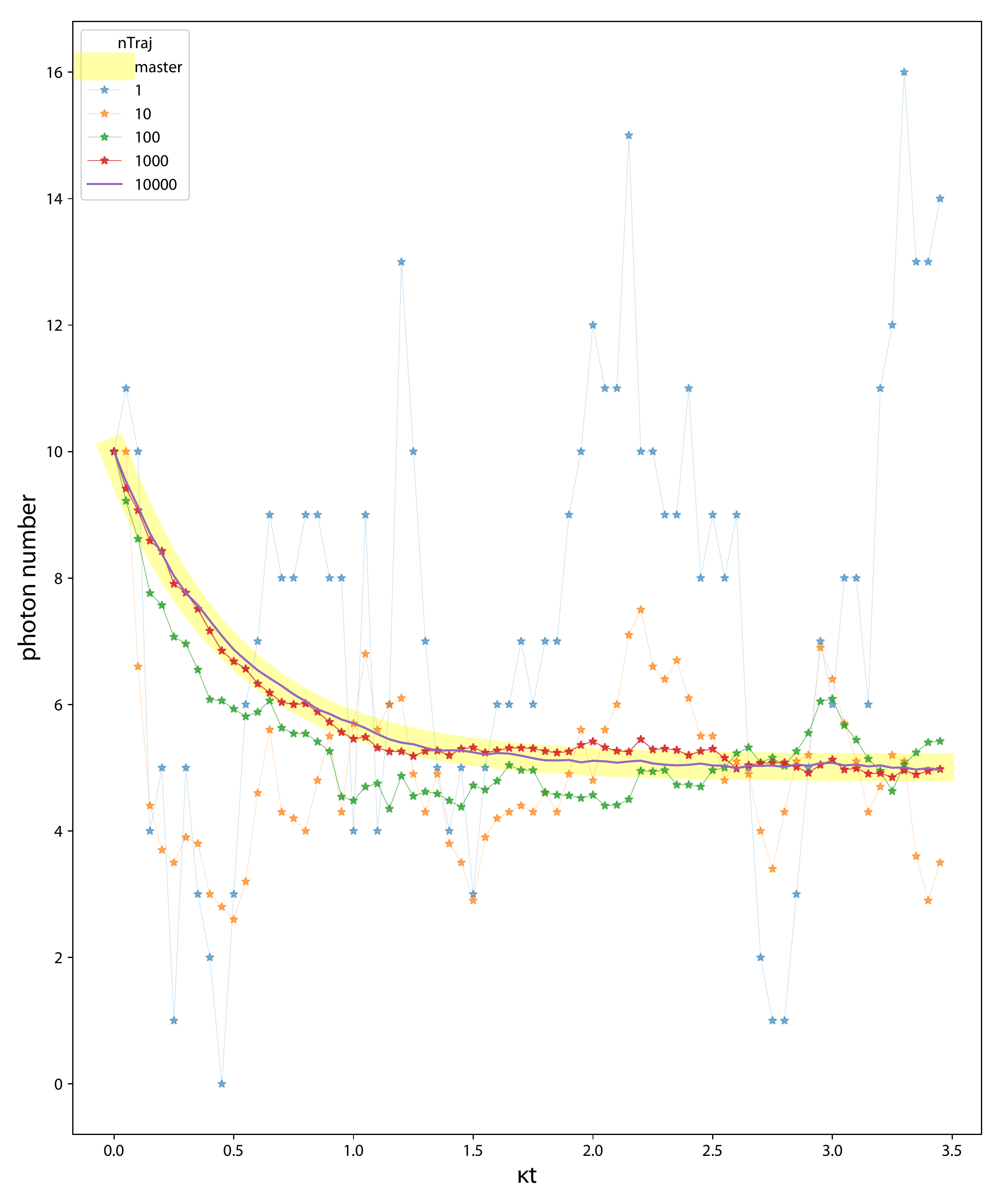}
 \caption{A typical picture of convergence of the MCWF solution to that of the master equation (midline of the yellow stripe) as a function of the number of trajectories. A single trajectory fluctuates wildly, without any appreciable relaxation of the initial state to the steady one. Even with 1000 trajectories we see significant deviations, but the average of 10000 trajectories fits nicely. Parameters: \(\nTh=5\), \(\ket{\Psi(0)}=\ket{10}\), \(\Dp=0.1\), \(\Delta t=0.05\).}
 \label{fig:exampleTrajs}
\end{figure}

Let us first study the case of \(\eta=0\) and Fock-state initial condition, which is the most demanding of simple examples. One reason for this is that in the lack of coherent driving, the just discussed stabilization mechanism on a single trajectory is absent. The other reason is a manifestation of bosonic enhancement: the photon exchange with the reservoir is the more intense, the larger the photon number the mode already has. Therefore, the photon number along a single trajectory fluctuates wildly, with cascading absorptions and emissions. To illustrate this, and give the reader a taste about the number of trajectories needed for acceptable convergence, we display the time evolution of the average photon number over trajectory ensembles of various sizes in \cref{fig:exampleTrajs}. Here, \(\nTh=5\) and \(\ket{\Psi(0)}=\ket{10}\), so that the photon number along the master-equation evolution stays in the interval between 5 and 10. Yet, in ensembles on the order of one million trajectories, we have encountered single trajectories that overshoot a Fock-state cutoff of 200!

The lack of driving and the Fock-state initial condition means that the MCWF method is drastically simplified to a classical Markov chain in Fock space, since in this case the mode state will remain a number state throughout the evolution. The process could be treated exactly by using random numbers with exponential distribution for the waiting time till the next jump, and simply hopping from one jump to the next in time. However, when regarded as a special case of our adaptive algorithm with the superimposed criterion of equally distributed sampling times, the behavior becomes nontrivial.

\begin{figure}
 \includegraphics[width=\linewidth]{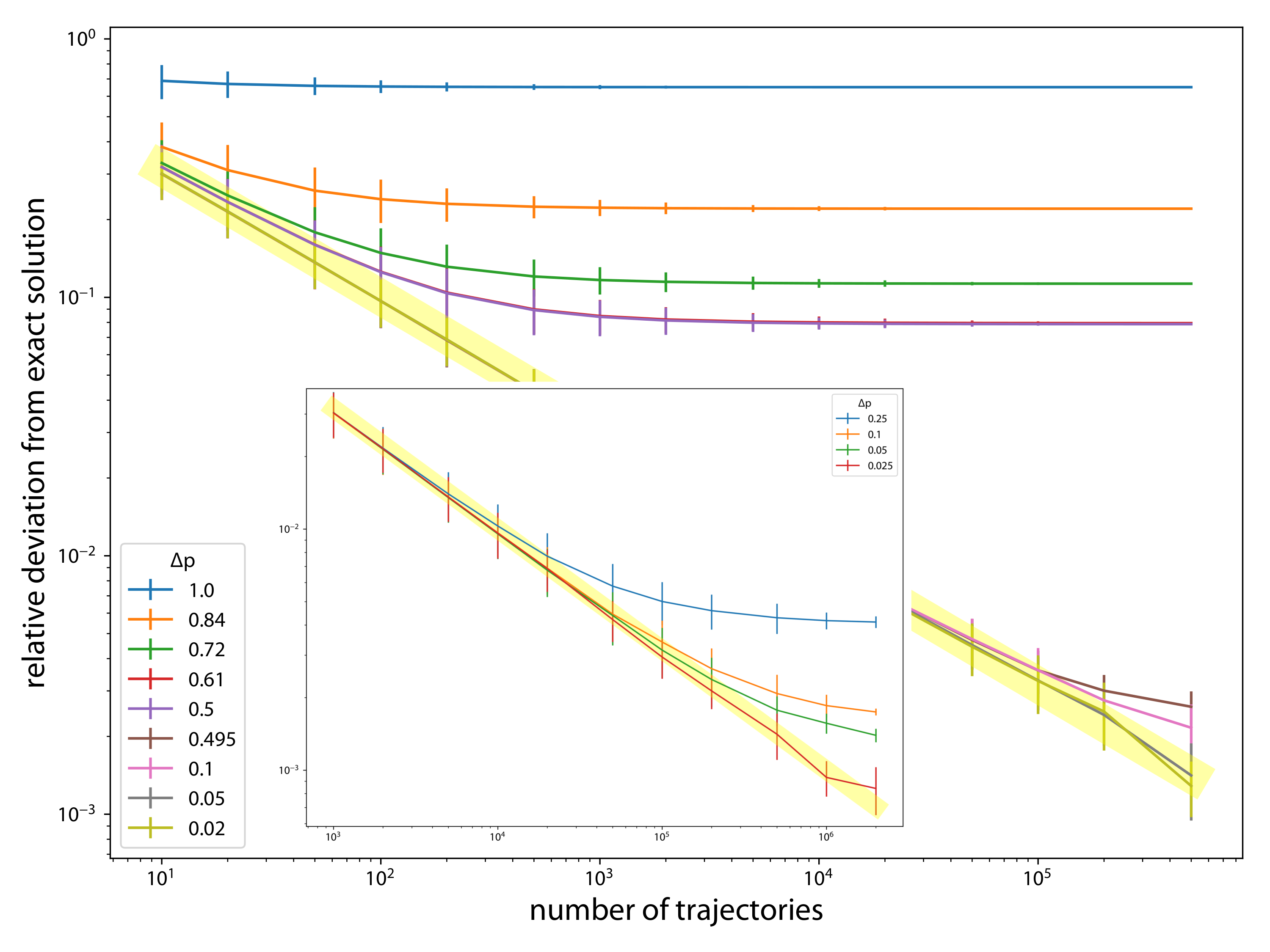}
 \caption{Same parameters as in \cref{fig:exampleTrajs}, but with varying \Dp. The midline of the yellow stripe is the line \(\frac1{\sqrt{\text{number of trajectories}}}\), which coincides with a linear fit (in log-log scale) on the curve for \(\Dp=0.02\). The three features to be noted in the Figure are: (1) the general trend of bettered convergence with decreasing \Dp; (2) the flattening out of the curves with increasing number of trajectories, which occurs at the larger number of trajectories, the smaller \Dp we have; (3) the critical behavior at \(\Dpc=0.5\) (the curve for \(\Dp=0.61\) virtually coincides with the one at 0.5!), below which there is a sharp drop in the deviation. The inset exposes the flattening-out behavior on a more suitable scale. These three features are explained in the text. To avoid misunderstandings, we note that here and throughout the paper, we use artificially big values of \Dp in order to magnify the effect of finite \Dp for the purpose of presentation. In practice, we use values on the order of \(10^{-2}\).}
 \label{fig:5Main}
\end{figure}

Our main result is displayed in \cref{fig:5Main}. Here we plot the deviation of trajectory ensembles from the exact (master-equation) solution as a function of the size of the ensemble. The deviation is measured by 
\begin{equation}
 \text{deviation}(f,g)\equiv\frac{2\norm{f-g}}{\norm{\abs{f}+\abs{g}}},\qqtext{with}
 \norm{f}\equiv\int_0^{T}\dd{t}\,\abs{f(t)},
\end{equation}
where \(f\) and \(g\) are functions of time over the interval \(\qty[0,T]\), these are the expectation values of the physical quantities along the trajectory.

The general trend that a decreasing \Dp betters the convergence is obvious, although we observe a striking critical behavior at \(\Dpc=0.5\): for larger \Dp values, the curve flattens out to a relatively high value with increasing number of trajectories and the curve already at \(\Dp=0.61\) is virtually indistinguishable from the limiting curve at \Dpc. On the other hand, for values smaller than the critical, the \Dp-dependence can be captured only with a number of trajectories on the order of hundreds of thousands: the larger \Dp, the smaller the number of trajectories for which the curves start to flatten out.

The MCWF method has two layers of errors. The first layer comes from discretization: the jumps can happen only at the endpoints of the timesteps. It is easy to see that the probability of a single jump happening evaluated at the end of the timestep is always smaller than what would come from the exact exponential waiting-time distribution, so that due to discretization the number of jumps gets smaller than the exact value. The second layer comes from the missed multi-jump events due to the first-order nature of the method, which also amounts to a smaller number of jumps than what we would have on an exact jump trajectory. As exposed in \Cref{app:Quantification}, both these kinds of error scale with the square of an average timestep, which in turn scales with \(\Delta p^2\).

The fact that the lower \Dp-curves in \cref{fig:5Main} follow a line \(\frac{\text{const.}}{\sqrt{\text{number of trajectories}}}\) (where the constant happens to be 1 within the errorbars of a least-square fit) up to a certain limiting number of trajectories, is a manifestation of the law of large numbers, given that the trajectories are independent. So what we observe in this region is the statistical error of averaging a finite set of independent trajectories. The flattening-out as a function of the trajectory number starts when this statistical error reaches the same order of magnitude as the intrinsic error of the method explained above, which of course happens with the larger number of trajectories, the smaller \Dp we use. However, even though our statistics is large, we do not seem to be able to verify the \(\Delta p^2\)-dependence of the intrinsic error in the figure.

\begin{figure}
 \includegraphics[width=\linewidth]{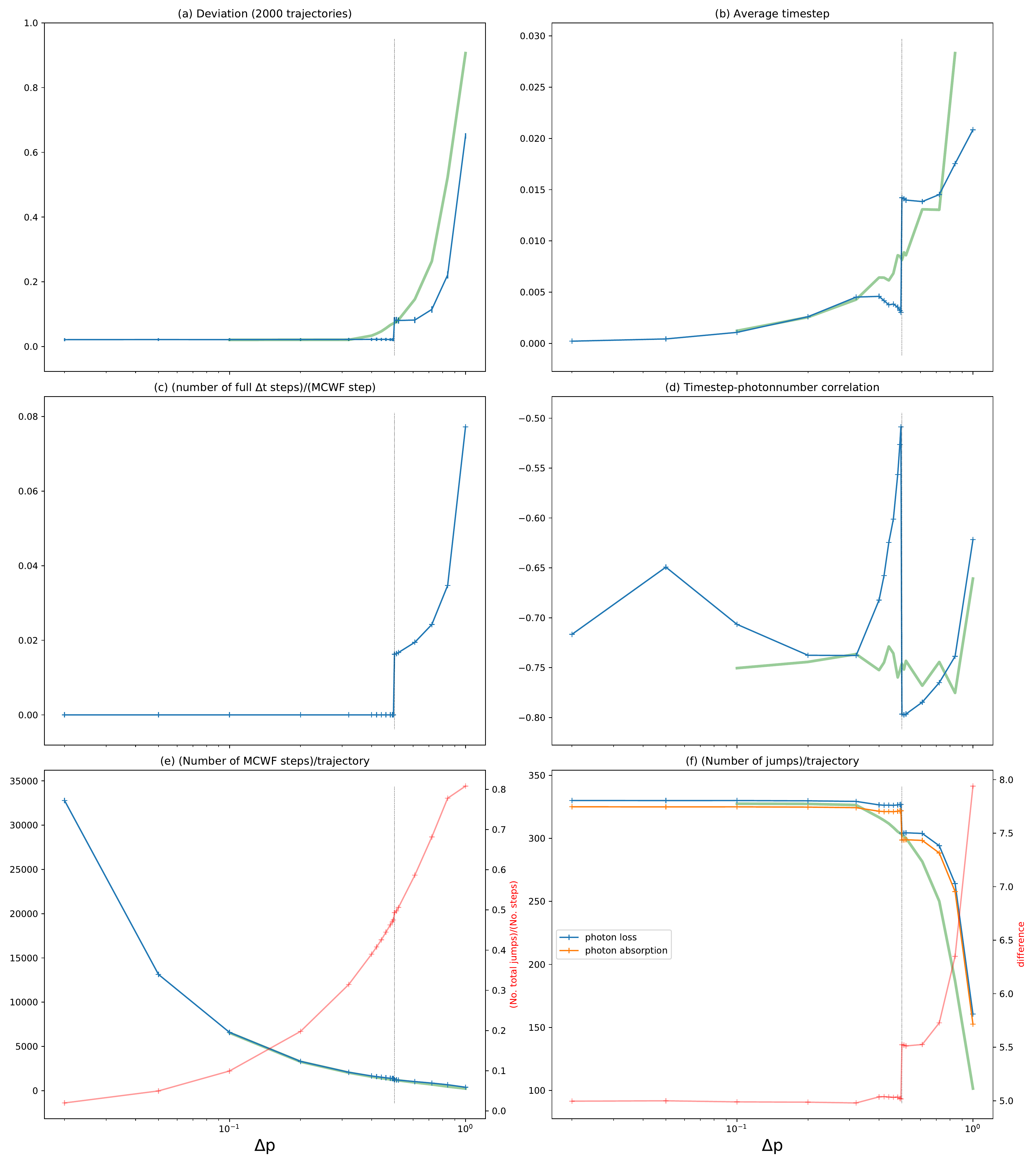}
 \caption{Important characteristics of MCWF evolution as a function of \Dp with same parameters as in \cref{fig:5Main}. The discontinuity observed on the panels is explained in detail in the text. For the green lines, \(\Delta t=0.25\), and they are plotted in order to show the behavior without discontinuity. In panel (f), the green line represent the mean of the two kinds of jumps. The average timestep on panel (b) and the correlation between the stepsize and the photon number in panel (d) are taken along a trajectory and over the full ensemble of trajectories as well.}
 \label{fig:5Logs}
\end{figure}

Let us return to the critical behavior, which is further exposed in \cref{fig:5Logs}, where we plot certain average characteristics of the trajectories as a function of \Dp, the criticality being present in each. On panel (f), we can see that for \(\Dp>\Dpc\), the method even misses the correct number of average jumps, and even the difference between the two kinds of jumps is rendered incorrectly. (This latter is obtained trivially: since the system starts form the 10-photon state, and the steady state is the 5-photon state [\(=n_\text{Th}\)], the average difference should be 5.) It should be noted that the number of MCWF steps continues its steep increase for subcritical \Dp values (cf. panel [e]), while the quality of convergence doesn’t increase appreciably (cf. panel [a]).

The “smoking-gun evidence” for the cause of the critical behavior is presented by the green lines on all panels and panels (b,c):
\begin{description}
 \item[The green lines] While in \cref{fig:5Logs}, the sampling time \(\Delta t=0.05\), for the green lines we chose \(\Delta t=0.25\), and see that the criticality disappears with this choice.
 \item[Panel (b)] The timestep averaged over the trajectory drops sharply at the critical point.
 \item[Panel (c)] The possibility of taking full \(\Delta t\) steps ceases at the critical point (curve drops to exactly 0 for \(\Dp<\Dpc\)).
\end{description}
In conclusion, the criticality depends strongly on \(\Delta t\), being related with the possibility of taking full \(\Delta t\) steps. Let us explain: In our present case, when the timestep is controlled purely by the relation \cref{eq:DpControl}, we can determine the condition for the system to be able to take a full \(\Delta t\) step when in the Fock state \(n\):
\begin{equation}
 \Delta t\leq\frac\Dp{r_\text{tot}(n)}=\frac\Dp{2\kappa\qty[\qty(\nTh+1)\,n+\nTh\qty(n+1)]}=\frac\Dp{2\kappa\qty[\qty(2\nTh+1)\,n+\nTh]}.
\end{equation}
This gives a critical \Dp value for each Fock state for the case when the equality sets in:
\begin{equation}
\label{eq:criticalDp}
\Dpc(n)= 2\kappa\,\Delta t\,\qty[\qty(2\nTh+1)\,n+\nTh].
\end{equation}
The smallest \Dp value given by this is for \(n=0\), which gives us the value of the critical point observed in the figure as:
\begin{equation}
 \Dpc=\Dpc(0)=2\kappa\,\nTh\,\Delta t,
\end{equation}
equaling 0.5 for the parameters used in \cref{fig:exampleTrajs,fig:5Main,fig:5Logs}. Why does the cessation of the possibility of taking full \(\Delta t\) steps cause a drop in the average timestep and a feature also in the other characteristics plotted in \cref{fig:5Logs}? Assume the system has just undergone a sampling at time \(u\,\Delta t\) with some \(u\in\mathbb N\), and it is in the 0th Fock state at this instant. Then, for supercritical \Dp, it will directly jump to the next sampling time instant \(\qty(u+1)\Delta t\), while for a \Dp value just below \Dpc, it will take a step just short of this next sampling time instant, so that it needs to take a very small step to finish the full \(\Delta t\) interval. It is the appearance of such small fragmentary steps that make the average timestep drop at the critical point.

The drop in the deviation of an ensemble of trajectories from the master-equation solution can also be explained from the drop seen in panel (b). In fact, the quality of convergence of the MCWF depends not directly on \Dp, but on the average stepsize: the smaller the timestep, the less probable we miss jumps via two-jumps-per-timestep events, hence, the better our (first-order) MCWF is. The dependence on \Dp is only via the dependence on the timestep.

The sharp feature in panel (d) at \Dpc can also be explained. Generally, the higher the photon number, the smaller steps we have to take, so that overall, the timestep-photonnumber correlation must be negative, which is what we see. The lower the \Dp, the stricter the stepsize control, so that the correlation with the photon number should increase in modulus, that is actually the case for \(\Dp>\Dpc\). At the critical value, the small fragmentary steps appear, whose size is largely independent of the photon number, hence the sudden drop in the modulus of the correlation.

\begin{figure}
 \includegraphics[width=\linewidth]{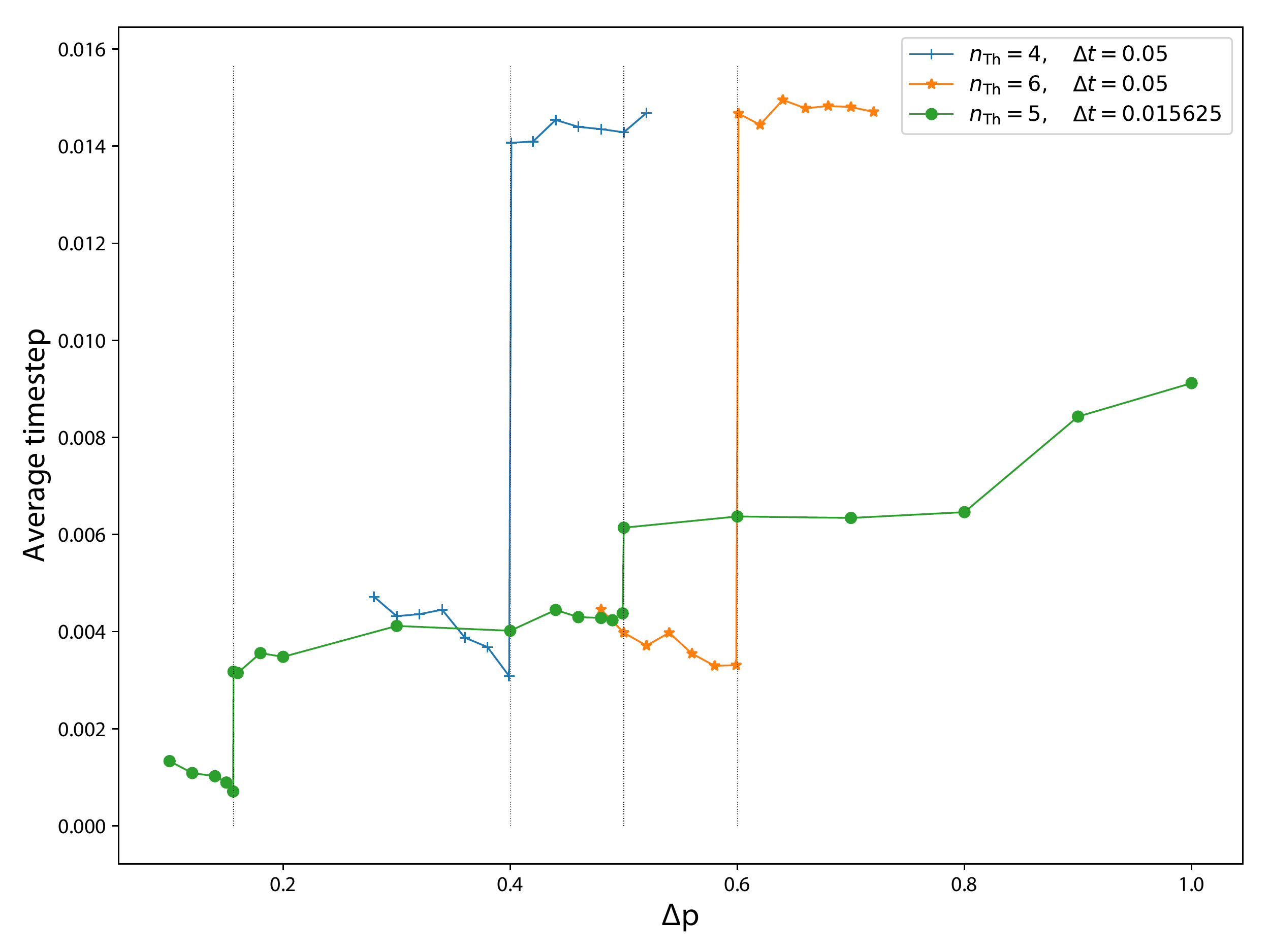}
 \caption{Behavior of the stepsize as a function of \Dp for three sets of parameters. The criticality (which is even doubled for the green line) follows the prediction of \cref{eq:criticalDp}.}
 \label{fig:timestepJumps}
\end{figure}

On changing parameters, the above picture of the criticality is confirmed. First of all, the green lines in \cref{fig:5Logs} represent the case \(\Delta t=0.25\), where the criticality disappears because \(\Dpc>1\) according to \cref{eq:criticalDp}. In \cref{fig:timestepJumps} we present three further cases: for \(\Delta t=0.05\) as above, but \(\nTh=4\) and 6 the critical point is given as 0.4 and 0.6; while in the case of \(\Delta t=0.015625\) and \(\nTh=5\), \cref{eq:criticalDp} even predicts two critical points below 1, since \(\Dpc(0)=0.15625\) and \(\Dpc(1)=0.5\). These predictions are confirmed by the figure.

\subsection{Contention with ODE-control}
\label{sec:contention}

Let us look at how the above picture is modified if the mode is driven coherently with Rabi frequency \(\eta\), that is, the Hamiltonian \labelcref{eq:modeHamiltonian} is nontrivial, meaning a nontrivial ODE evolution with its own internal stepsize control. This control will contend with our superimposed \Dp-control.\footnote{\label{note:ODE}For the sake of clarity, we remark that \emph{a priori,} jumps never come without a Hamiltonian to reckon with in the form of the non-Hermitian part, cf. \cref{eq:nHH}. Furthermore, this part scales with system parameters and state in the same way as the total jump probability. In the case when this “obligatory” part of \HnH\ is diagonal in the working basis, it can be treated with an exact propagator (this is what C++QED does), so that it does not burden the ODE stepper. In the case studied in \cref{sec:pureDpControl}, when the state remains a Fock state all along, this term moreover amounts to nothing more than a trivial norm factor, which anyway disappears during the renormalization after each timestep. So in this case it is possible to completely disregard this “coherent” part of the evolution. In more involved uses, when the working basis cannot be chosen in such a way that the non-Hermitian part be diagonal, the ODE-control will dominate the timestep control due to the just mentioned scaling argument, and the superimposed \Dp-control will intervene only at exceptional moments. The same is true when other parts of the Hamiltonian have the largest characteristic frequencies.}

In many situations of physical interest, a generic behavior that was noticed already in \cite{Vukics2005a} is that off-diagonal elements of the density matrix converge slower than diagonal ones. In the present case this makes that the phase of the field converges worse than the amplitude, which we have found difficult to prove in a clear-cut way. A possible physical interpretation of this behavior could be that the photon loss measures the photon number. This suggests that a homodyne detection could result in opposite behavior, however, this we could not prove either in a distilled way.

\begin{figure}
\begin{center}
\begin{tabular}{l r}
 (a) & \includegraphics[width=.8\linewidth]{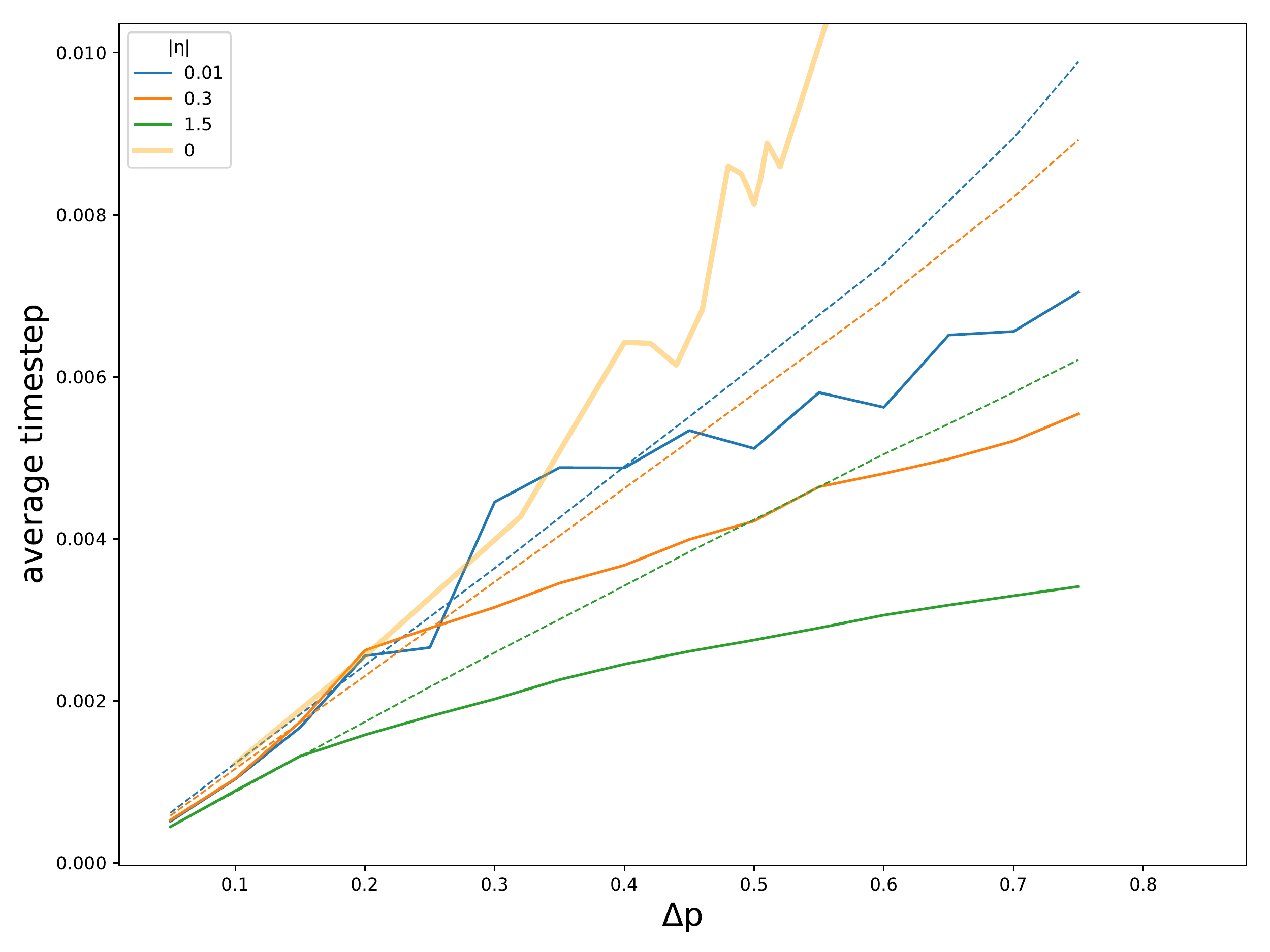}\\
 (b) & \includegraphics[width=.8\linewidth]{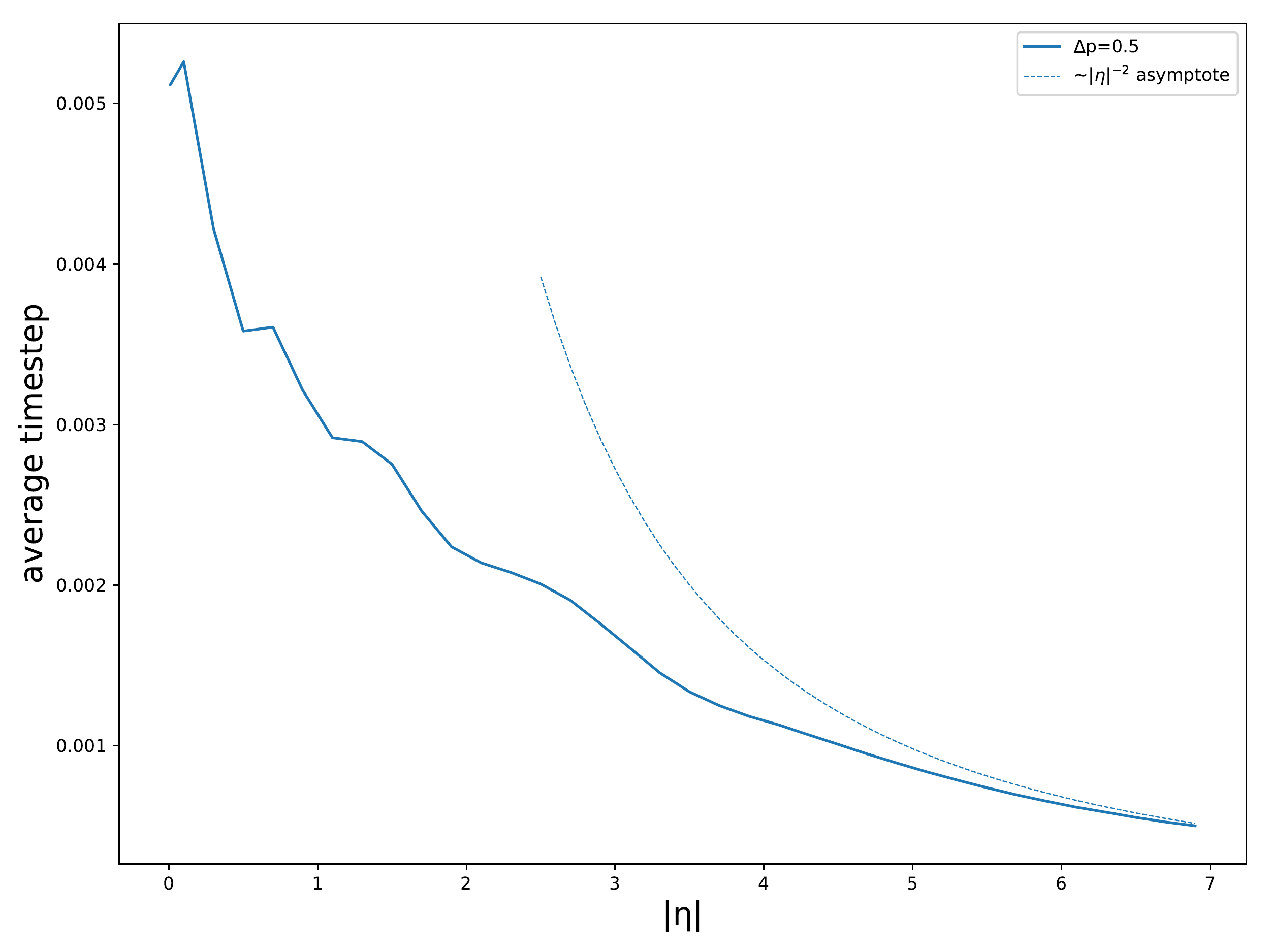}
\end{tabular}
\end{center}
 \caption{The behavior of the algorithm as witnessed by the average timestep in the case of nontrivial coherent evolution. Same physical system and parameters as in \cref{fig:5Main}, but with finite driving amplitude \(\eta\). In panel (a), the dashed lines are predictions from \cref{eq:dpControlledTimestep}, with the same color code as for the solid lines.}
 \label{fig:eta_dp_scan}
\end{figure}

In \cref{fig:eta_dp_scan}(a) we display the average timestep as a function of \Dp with three different values of \(\eta\). The dashed lines are predictions based on \cref{eq:DpControl} assuming pure \Dp-control:
\begin{equation}
\label{eq:dpControlledTimestep}
 \overline{\delta t}_\text{\Dp-controlled}=\Dp\,\overline{\frac1{r_\text{tot}}}=\Dp\,\overline{\frac1{2\kappa\qty(\qty(2\nTh+1)a^\dagger a+\nTh)}},
\end{equation}
where the overline means averaging over time along one trajectory and averaging over (an ensemble of 20000) trajectories as well.

The main message of the figure is that for small \Dp, \Dp-control dominates, so that the curves overlap with \Dp-controlled timestep, while for increasing \Dp, ODE-control takes over, so that the curves flatten out, the timestep becoming independent of \Dp. ODE-control depends on the largest frequency present in the system, the resulting stepsize scaling with the inverse of this frequency. This means that this control is the stricter, the larger the frequencies present in the system. This is the reason why the larger the \(\eta\), the lower the \Dp value at which the curves start to flatten out. For the same reason, the stepsizes are generally smaller for increasing \(\eta\).

For large enough \(\eta\) values, pure ODE-control is established, as exhibited in panel (b) of the same figure. Here we plot the dependence of the timestep on \(\abs{\eta}\) for a rather large \Dp value. We observe that the curve asymptotically coincides with the dashed one which represents a \(\propto\abs{\eta}^{-2}\) decrease. The reason for this is that in this simple case, the largest frequency in the system scales with \(\eta^2\), since in the Hamiltonian \labelcref{eq:modeHamiltonian}, the mode amplitude is also proportional to \(\eta\).

\section{Comparison with the integrating method of MCWF evolution}
\label{sec:outlook}
\begin{figure}
 \centerline{\includegraphics[width=.75\linewidth]{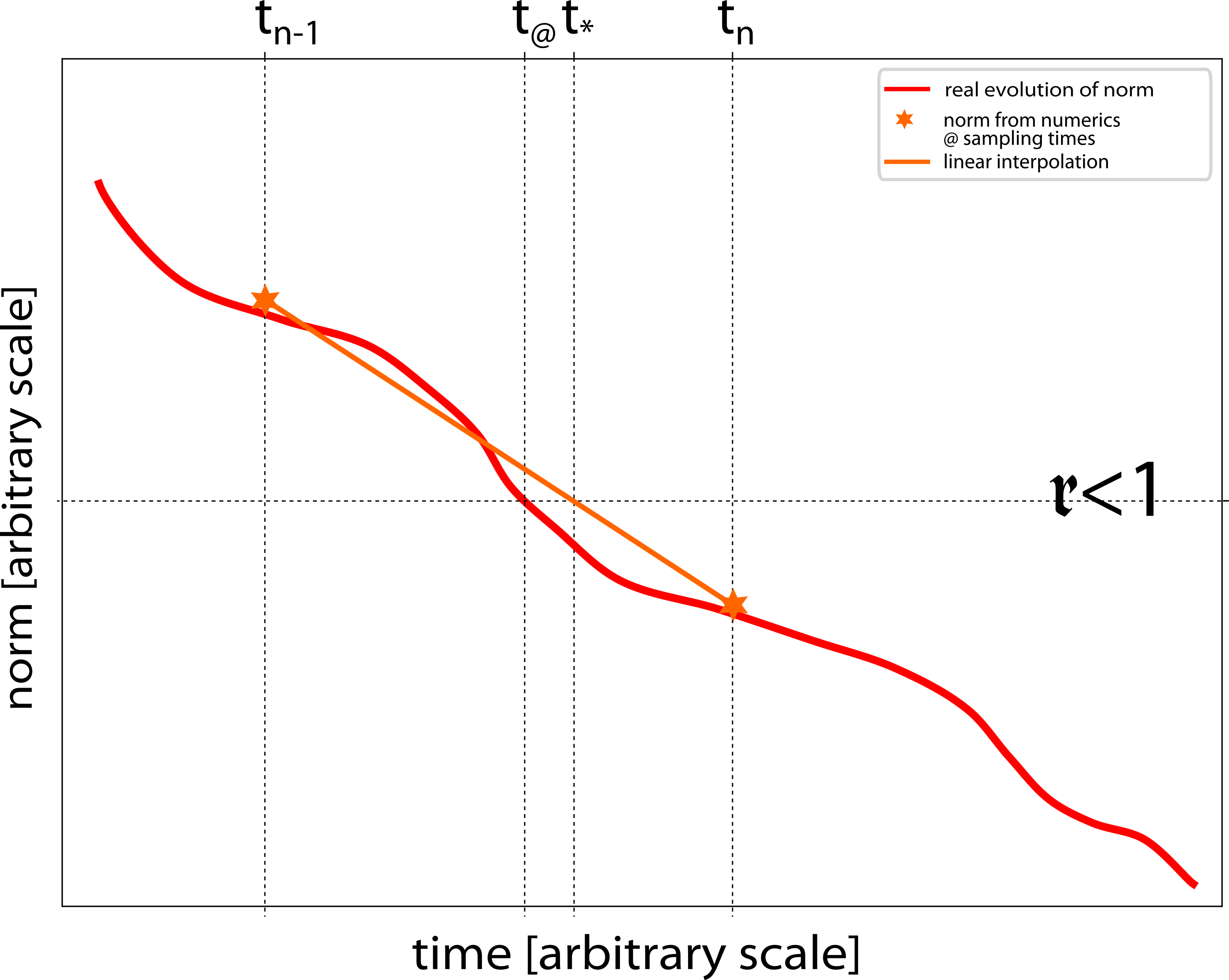}}
 \caption{Cartoon illustrating the workings of the integrating method. \(t_n\)s are the sampling time instants when the state-vector norm is compared against the previously drawn random number \(\mathfrak{r}\). \(t_@\) would be the time instant of the jump according to the real evolution of the norm, while \(t_*\) is the jump time instant retrieved by the algorithm when using the most primitive retrieval method: linear interpolation.}
 \label{fig:integrating}
\end{figure}

In the implementation of the MCWF method, there is another main stream, which we will refer to as “integrating” in contrast to our (adaptive) “stepwise” method. In this algorithm, the norm of the state vector is let to decrease under the evolution by \HnH. A random number \(\mathfrak r\in[0,1)\) is drawn at the beginning, and as the state-vector norm reaches this number, a jump is introduced. At this point, the distribution of jumps \(p_m\) is calculated, and the \(n\){}th jump is performed where $n$ is the smallest integer satisfying \(\sum_{i=1}^n p_i\geq\mathfrak r\). It can be shown that the norm-loss equals the accumulated jump probability (hence the name “integrating method”) under a very general set of conditions. The workings of the method together with the typical error that it involves are illustrated in \cref{fig:integrating}. One needs to define a set of sampling times \(t_n\) where the norm will be compared against \(\mathfrak{r}\). It is a non-trivial issue what is a good sampling interval. The source of the method’s error is that when we notice that the norm has shrunken below \(\mathfrak{r}\) at time \(t_n\), then we are already after the real time instant of the jump. Therefore, we need a mechanism to retrieve the jump time instant together with the state of the system at that time in order to perform the jump. When using linear interpolation, the error will be \(\abs{t_@-t_*}\), which somehow scales with the sampling time interval.

One of the advantages of the method is that it enables the use of multistep methods for the ODE evolution part, although the multistepper has to be exited now and again to check the norm of the state vector, and eventually retrieve the jump time instant and perform the jumps.

The \texttt{mcsolve} routine of QuTiP \cite{Johansson2012,Johansson2013}, uses the integrating method, the jump time within a supplied norm tolerance being retrieved by bisection root-finding combined with linearization (Paul Nation, private communication). The parameters of this algorithm are the norm tolerance (default: \(0.001\)) and the number of maximum iterations of the root-finding algorithm (default: \(5\)).

Our stepwise method is more conservative, conceptually simpler and more robust in the sense that it does not rely on a heuristic for retrieving the jump time instance, hence it is immune against failures of such a heuristic (for which there are specific error messages in QuTiP).\footnote{According to the developers (Paul Nation, private communication), in all the usecases encountered so far, the jump time was found within the default tolerance in at most 3 iterations.} However this comes at the price of a certain reduction of performance, which is twofold:
\begin{enumerate}
 \item The stepsize is bounded in each timestep due to the \Dp-criterion, while in the integrating method it is solely the ODE stepper which controls the stepsize. This difference, however, disappears in the case when the ODE-control dominates the timestep-control, which in our experience is the case in most real-life situations (cf. \cref{note:ODE}).
 \item The jump probabilities have to be evaluated in each timestep, instead of just calculating the norm. This overhead on the other hand can become negligible if the evaluation of the Hamiltonian (which is done several times per step within the ODE solver) is significantly more expensive, which is the case in most real-life examples we encountered so far.
\end{enumerate}

Let us make two more notes of comparison favoring our method.
\begin{enumerate}
 \item The integrating method requires more parameters for controlling the error of the MCWF, since besides the norm tolerance, further parameters are required for the method dedicated to retrieve the time instance of the jump (e.g. number of iterations). Furthermore, the parameters controlling the specific error of the first-order MCWF are intertwined with the parameters controlling the sampling, since the larger the sampling intervals, the larger the overshoot of the real jump time instant. This is in contrast to our single parameter \Dp, and our stepsize control which is done either by the ODE stepper, or by the \Dp condition, depending on which one is stricter. In our method it is also easy to estimate the maximum probability of two jumps occuring in a single time step, since assuming \(\Dp\ll1\), this is given by \(\Dp^2\).
 \item The possibility of exactly renormalizing the state vector after each ODE step is lost in the integrating method. This is an important stabilizing means of the method (cf. \cref{sec:favoringUsecases}), which is available in our algorithm.\footnote{On a final note: the concept of a single adaptive step is well-defined in our case due to the possibility of jumps being immediately accounted for in each step, making our algorithm compatible with higher-level trajectory drivers.}
\end{enumerate}

\begin{figure}
 \includegraphics[width=\linewidth]{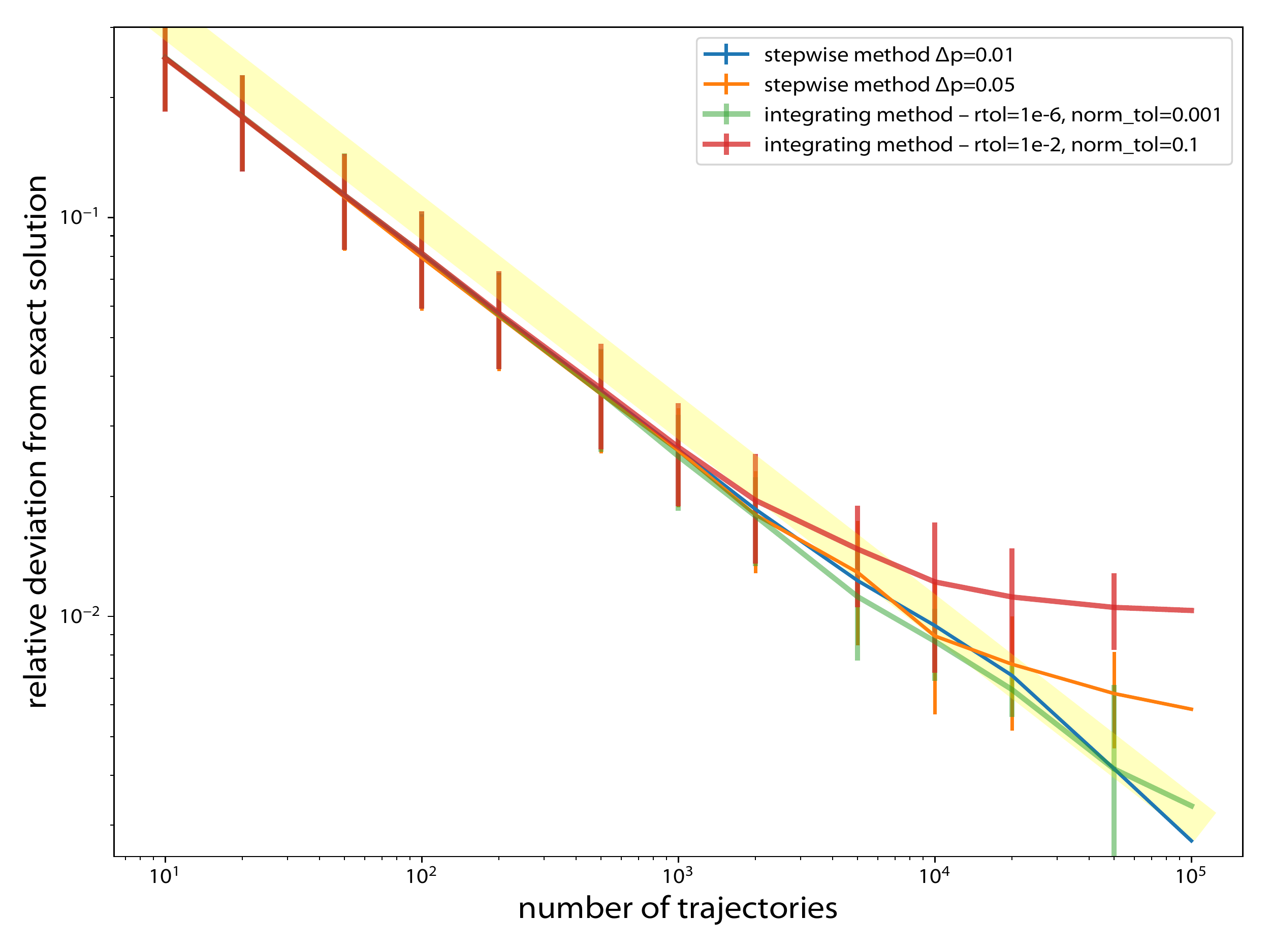}
 \caption{Comparison of the convergence of our adaptive stepwise algorithm with two different \Dp{s} and the integrating algorithm as implemented in QuTiP with two different sets of precision parameters. The physical system is the same as in \cref{sec:contention}, with same parameters as in \cref{fig:eta_dp_scan}. The yellow stripe is the same as in \cref{fig:5Main}.}
 \label{fig:qutip}
\end{figure}

Having said all this, the convergence properties of the two methods are similar when respective appropriate parameters are chosen, as illustrated on the example of the coherently driven mode interacting with a thermal bath in \cref{fig:qutip}. As we see, the “flattening out” behavior also appears in the case of the integrating method. Note that this method is also sensitive to the issue of double jumps, being also first order in this sense: it will miss such events when two jumps would occur within the time interval corresponding to the given norm tolerance.

\subsection{Example usecases favoring the stepwise method}
\label{sec:favoringUsecases}
\paragraph{Moving-particle cavity QED}
Let us consider a particle of mass \(m\) with a one-dimensional motional degree of freedom. We consider periodic boundary condition in space, meaning that the particle momentum is discretized with intervals \(\Delta k\), so that it is possible to define a dimensionless “wave-number operator” for the particle \(\mathbb{K}=p/(\hbar\,\Delta k)\), that has integer spectrum. The particle is moving in a single-cosine-mode optical field with wave number \(K\) that is an integer multiple of \(\Delta k\). The Hamiltonian then reads:
\begin{equation}
 H=\hbar\omega_\text{rec}\,\mathbb{K}^2+V\cos^2\qty(Kx)
 ,
\end{equation}
where the recoil frequency is defined as \(\omega_\text{rec}=\hbar\,\Delta k^2/(2m)\), and \(V\) is an energy scale representing the coupling between the mode and the particle. This is a numerically demanding problem because of the quadratic increase of the characteristic frequencies of the different components particle wave-number space. The quadratic increase makes that very separate timescales appear in the simulation, resulting in very small timesteps compared to the necessary simulation time. (In general, the simulation time scales with the slowest, while the timestep with the fastest timescale.) This situation can be improved by transforming into an interaction picture defined by the kinetic part of the Hamiltonian:
\begin{equation}
 H_\text{I}=\frac V2\qty[e^{4i\omega_\text{rec}\qty(\mathbb{K}-K/\Delta k)t} e^{-2iKx}+e^{-4i\omega_\text{rec}\qty(\mathbb{K}+K/\Delta k)t} e^{2iKx}],
\end{equation}
the gain by this being that the characteristic frequency now increases only linearly with \(\mathbb{K}\). This can lead to an increase by a few orders of magnitude in the timestep. However, the still large frequencies in the now explicitly time-dependent Hamiltonian can lead to instabilities in the ODE stepper. We have found that an exact renormalization of the state vector after each ODE step resolves this issue. This approach has been used with success in several situations \cite{Vukics2005a,Vukics2005b,Vukics2007,Vukics2009,Schulze2010,Niedenzu2010,Niedenzu2012,Sandner2013,Winterauer2015}.

\paragraph{Non-unitary interaction picture}
In many situations it is worthwhile to use not only a traditional interaction picture, but an exact propagator obtained by exponentializing the full diagonal part of the non-Hermitian Hamiltonian, that is, a non-unitary interaction picture. Consider \cref{eq:HnH_mode}, but with off-resonant driving with detuning \(\delta\):
\begin{equation}
 \HnH=-\hbar\qty[i\kappa\qty(2\nTh+1)+\delta]a^\dagger a+i\hbar\,\eta\,\qty(a^\dagger-a).
\end{equation}
Here, if we pass to a traditional interaction picture
\begin{equation}
 H_\text{nH,I}=-\hbar i\kappa\qty(2\nTh+1)a^\dagger a+i\hbar\,\eta\,\qty(e^{-i\delta t}\,a^\dagger-e^{i\delta t}\,a),
\end{equation}
then, a term contributing high frequencies with growing excitation number remains in the form of the non-Hermitian term. It is much better to eliminate this as well:
\begin{equation}
 H_\text{nH,nU}=i\hbar\,\eta\,\qty(e^{\qty[\kappa\qty(2\nTh+1)-i\delta]t}\,a^\dagger-e^{\qty[-\kappa\qty(2\nTh+1)+i\delta]t}\,a).
\end{equation}
We see that the frequency depending linearly on the excitation number has disappeared, which again can lead to a substantial increase in the timestep. The downside is that there appeared explicitly time-dependent terms, some of which grow while others decrease exponentially in time. This can again lead to instabilities in the ODE stepper. Here again, we have found that these instabilities are removed by an exact renormalization of the state vector after each ODE step. This approach has been used with success in several situations \cite{Dombi2013,Dombi2015,Fink2017}.

The necessity of renormalization after each ODE step in the two situations shown in this section has the consequence that the integrating method cannot be used, since the norm remains 1 during the whole evolution.

\section{A note on sampling and time averaging}
\label{sec:sampling}
Because of the adaptive nature of the trajectories, there are two possibilities for sampling along the evolution of a single trajectory: one can either sample (1) in equal time intervals or (2) in equal number of steps. Sampling method (2) is better suited for following the physics of the problem along a single trajectory, since at those times where a lot of dynamics takes place, the stepsize control will choose smaller timesteps resulting in more samples than in calmer times. Moreover, it is only with method (2) that the sampling does not influence the trajectory simulation in the way we saw in \cref{sec:pureDpControl}.

In this connection, it is of great importance to note that when using a long single trajectory for finding steady-state results as time averages, then with sampling method (2) the time average must be calculated with weighing with the stepsize:
\begin{subequations}
\begin{equation}
 \overline{\mathcal O}^{(2)}=\frac{\sum_{m\in[\text{sampling steps}]}\delta t(m)\,\expval{\mathcal O}\qty(m)}{\sum_{m\in[\text{sampling steps}]}\delta t(m)},
\end{equation}
where \(\mathcal O\) is an observable and \(\delta t(m)\) is the timestep done in the \(m\){}th step.
This is because states are correlated with stepsize (cf. \cref{fig:5Logs}(d)) and hence the density of samples, so that states resulting in smaller stepsize (e.g., states with higher photon numbers in the example of \cref{sec:pureDpControl}) will be overrepresented among the samples. On the other hand, with sampling method (1), stepsize-weighing must not be used:
\begin{equation}
 \overline{\mathcal O}^{(1)}=\frac{\sum_{t_i\in[\text{sampling times}]}\expval{\mathcal O}\qty(t_i)}{\qty(\text{number of sampling times})},
\end{equation}
\end{subequations}
where the sampling times are equally distributed in time. Confusion in this respect can result in gross misestimates of steady-state values!

\section{Conclusion}
Besides presenting a stepwise adaptive MCWF algorithm controlled by \Dp, the total jump probability per timestep, in this paper we have studied the convergence behavior of the MCWF method depending on this parameter. This dependence should be equivalent with the dependence on the norm-tolerance parameter in the case of that implementation of the MCWF method which we came to call “integrating”, and should pertain to any other implementation of the MCWF method as well.

We have found that the \(\propto\frac1{\text{number of trajectories}}\) dependence of the deviation of the MCWF from the exact solution flattens out, which happens at the larger number of trajectories, the smaller the \Dp. This behavior we attributed to the inherent errors of the first order MCWF method, namely (1) time discretization and (2) missing multi-jump events, both of which are \(O(\delta t^2)\) errors, meaning that they scale as \(\Dp^2\).

We have found and characterized a discontinuous behavior of the method as a function of \Dp\ in the pure \Dp-controlled case, which we have attributed to the way the trajectories are sampled in time. This a showcase of how sampling can modify the behavior of trajectories through influencing the stepsize, which at the same time displays the role of the average stepsize as the real regulator of MCWF convergence. 

In the case when a non-trivial Hamiltonian is present, we have characterized the contention between \Dp- and ODE-control, finding that ODE-control will take over when those characteristic frequencies of the system that do not participate in the loss increase. The takeover of ODE-control is signalled by that the average timestep becomes determined by the largest non-loss-related characteristic frequency of the system, becoming independent of \Dp. 

In the comparison with the integrating method, we have come to the conclusion that our stepwise method is more conservative, conceptually simpler, and more robust at the price of some reduction of performance. This reduction, however, should become marginal in many real-life situations. Regarding convergence, the two methods are equivalent when the respective parameters are chosen appropriately. In some special but realistic situations when a renormalization of the state vector relieves numerical problems of the ODE evolution, our method is favored above the integrating one.

\section*{Code availability}
\label{sec:code}
The algorithm presented here is available as the \texttt{quantumtrajectory::MCWF\_Trajectory} class in C++QED: a C++/Python application programming framework for simulating open quantum dynamics \cite{VukicsCppQEDa,VukicsCppQEDb,Sandner2014CppQED}.\footnote{For the development version and an extensive documentation, cf. \url{http://cppqed.sf.net}.} In particular, all the simulations presented here can be reproduced using the \texttt{PumpedLossyMode\_C++QED} script available in the framework’s distribution. A sample command line simulating a single trajectory may read:
{\small\[\text{\texttt{PumpedLossyMode\_C++QED --dpLimit 0.1 --seed 1000 --cutoff 2000 --nTh 5}}\]}
(the parameter \Dp is called \texttt{dpLimit} in the framework, for historical reasons). Some further code snippets are presented in \cref{app:pythonSnippets}.

\section*{Acknowledgments}
We acknowledge Peter Domokos, Helmut Ritsch, Sebastian Krämer, Raimar Sandner, Lajos Diósi, and Paul Nation for helpful discussions. On behalf of Project WAL, we thank for the usage of MTA Cloud (https://cloud.mta.hu/) that significantly helped us achieving the results published in this paper. This work was supported by the National Research, Development and Innovation Office of Hungary (NKFIH) within the Quantum Technology National Excellence Program (Project No. 2017-1.2.1-NKP-2017-00001) and Grants No. K115624 and K124351. A. V. acknowledges support from the János Bolyai Research Scholarship of the Hungarian Academy of Sciences.

\appendix
\section{Quantification of the error}
\label{app:Quantification}
In this appendix, we give a simple quantification of the behavior of the error in \cref{fig:5Main}.

Let us introduce some notation. Let $X(t)$ denote the photon-number process. In the simplest case exhibited in \cref{sec:pureDpControl}, when the mode is driven solely by the interaction with the thermal bath, $X(t)$ is a birth-death process with generator
\begin{equation}
q_{nm}= \begin{cases}
	 \lambda_n & \qqtext{if} m=n+1 \\ 
	 -\lambda_n-\mu_n & \qqtext{if} m=n \\
	 \mu_n & \qqtext{if} m=n-1 \\ 
	 0 & \qqtext{otherwise.}
\end{cases}
\end{equation}
where $\lambda_n= 2\kappa(n+1) n_{th} $ and $\mu_n=2\kappa (n_{th}+1) n. $ The theory of such processes is well developed, and it is known that the process $X$ is fully determined by its state space and its generator, furthermore one can easily determine the stationary state (if it exists), distribution of waiting times, probability of extinction etc. 

Let $Y(t)$ denote the discretized model. In this simple case $Y(t)$ is also a stochastic process, in particular a discrete-time Markov chain, but with different transition probabilities. Since the rate of transition is fixed between jumps, so is the timestep $\delta t$. Here the transition matrix is given by
\begin{equation}
p_{nm}=\expprob{P}{Y(t+\delta t)=m}{Y(t)=n}= \begin{cases} \lambda_n \delta t & \qqtext{if } m=n+1 \\ 1-(\lambda_n+\mu_n)\delta t & \qqtext{if } m=n \\ \mu_n \delta t & \qqtext{if } m=n-1 \\ 0 & \mbox{otherwise.}  \end{cases}
\end{equation}
Note that this matrix is a valid transition matrix if and only if $(\lambda_n+\mu_n)\,\delta t \leq 1$, furthermore that $Y(t)$ is a time-discretized version of $X(t)$. In order to understand the relaxation of the curves in the figures let us examine the quantity
\begin{equation}
\expprob E{X(t+\delta t) - Y(t+ \delta t)}{X(t)=Y(t)=n}, 
\end{equation}
that is, the difference of the mean of the real trajectories and that of the approximated ones at one timestep away from $t$, given that the trajectories coincide at time $t$. The expected value of $Y$ is given by
\begin{equation}
\expprob E{Y(t+\delta t)}{Y(t)=n}= n+\qty(\lambda_n-\mu_n)\,\delta t,
\end{equation}
while that of $X$ needs a little bit more effort to calculate.  
First we will show that
\begin{equation}
\expprob P{\abs{X(t+\delta t)-n}\geq 3}{X(t)=n}=O\qty(\delta t^3). 
\end{equation}
It is known that the waiting times between two jumps of a continuous-time Markov chain are exponentially distributed and independent random variables. Consider some independent, exponentially distributed random variables $T_j,j=1,2,3$ with parameters $ \gamma_j,j=1,2,3$. Then
\begin{multline}
	\mathbb P\qty(T_1+T_2+T_3< \delta t ) \\= \frac{\gamma_1\gamma_2\gamma_3}{\gamma_1-\gamma_2}\qty[ \frac{\frac{1}{\gamma_2}\qty(1-e^{-\gamma_2\delta t})- \frac{1}{\gamma_3}\qty(1-e^{-\gamma_3\delta t})}{\gamma_3-\gamma_2}- \frac{\frac{1}{\gamma_1}\qty(1-e^{-\gamma_1 \delta t})-\frac{1}{\gamma_3} \qty(1-e^{-\gamma_3\delta t})}{\gamma_3-\gamma_1} ]= O\qty(\delta t^3),
\end{multline}
which means that the probability of a continuous time Markov chain jumping at least three times in an interval of length $\delta t$ is $O\qty(\delta t^3)$, thus
\begin{multline}
\expprob E{X(t+\delta t)}{X(t)=n}= \sum_{m} m\,\expprob P{X(t+\delta t)=m}{X(t)=n}\\=\sum_{m\,:\,\abs{m-n}\leq2}m\,\expprob P{X(t+\delta t)=m}{X(t)=n} +O\qty(\delta t^3).
\end{multline}
Hence, for an at-most-second-order-in-$\delta t$ calculation of the expectation, we only need to calculate the probabilities 
\begin{equation}
\expprob P{X(t+\delta t)=m,\;\text{in at most 2 jumps}}{X(t)=n}, \qqtext{with}m=n,\;n\pm1,\;n\pm2. 
\end{equation}
These are found to read
\begin{subequations}
\begin{multline}
\expprob P{X(t+\delta t)=n,\;\text{\#jumps}\leq 2}{X(t)=n}\\= e^{-q_n\delta t}+\frac{\lambda_n\,\mu_{n+1}}{q_n-q_{n+1}}\qty(\frac{e^{-q_n\delta t}- e^{-q_{n+1}\delta t}}{q_{n}-q_{n+1}}- \delta t\, e^{-q_n\delta t})\\+ \frac{\mu_n\,\lambda_{n-1}}{q_n-q_{n-1}}\qty(\frac{e^{-q_n \delta t}- e^{-q_{n-1}\delta t}}{q_{n}-q_{n-1}}- \delta t\,e^{-q_n \delta t}) \\=1-q_n\delta t+\frac{(q_n\delta t)^2}{2}+\frac{\lambda_n\,\mu_{n+1}+\lambda_{n-1}\,\mu_n}2\delta t^2 + O\qty(\delta t^3),
\end{multline}
\begin{multline}
\expprob P{X(t+\delta t)=n+1,\;\text{\#jumps}\leq 2}{X(t)=n}\\=\frac{\lambda_n}{q_{n+1}-q_n}\qty(e^{-q_n \delta t}- e^{-q_{n+1}\delta t}) = \lambda_n \delta t -\frac{\lambda_n}2\qty(q_n+q_{n+1})\delta t^2 + O\qty(\delta t^3), 
\end{multline}
\begin{multline}
\expprob P{X(t+\delta t )=n-1,\;\text{\#jumps}\leq 2}{X(t)=n}\\=\frac{\mu_n}{q_{n-1}-q_n}\qty(e^{-q_n\delta t}-e^{-q_{n-1}\delta t})= \mu_n \delta t- \frac{\mu_n}2(q_n+q_{n-1})\delta t^2 +O\qty(\delta t^3),
\end{multline}
\begin{multline}
\expprob P{X(t+\delta t)=n+2,\;\text{\#jumps}\leq 2}{X(t)=n}\\=\frac{\lambda_n\lambda_{n+1}}{q_{n+2}-q_{n+1}}\qty(\frac{e^{-q_n\delta t}-e^{-q_{n+1}\delta t}}{q_{n+1}-q_n}- \frac{e^{-q_n \delta t}- e^{-q_{n+2}\delta t}}{q_{n+2}-q_n})=\frac12\lambda_n\lambda_{n+1} \delta t^2 + O\qty(\delta t^3),\\
\end{multline}
\begin{multline}
\expprob P{X(t+\delta t)=n-2,\;\text{\#jumps}\leq 2}{X(t)=n}\\=\frac{\mu_n\mu_{n-1}}{q_{n-2}-q_{n-1}}\qty(\frac{e^{-q_n\delta t}-e^{-q_{n-1}\delta t}}{q_{n-1}-q_n}- \frac{e^{-q_n \delta t}- e^{-q_{n-2}\delta t}}{q_{n-2}-q_n})= \frac12 \mu_n\mu_{n-1}\delta t^2 + O\qty(\delta t^3),
\end{multline}
\end{subequations}
from which we obtain
\begin{multline}
\expprob E{X(t+\delta t)-Y(t+\delta t)}{X(t)=Y(t)=n} \\= \frac{\delta t^2}{2}\bigg(q_n^2 + \lambda_n\mu_{n+1}+\lambda_{n-1}\mu_n + (n+1) \lambda_n(q_n+q_{n+1})\\+(n-1)\mu_n(q_n+q_{n-1})+(n+2)\lambda_n\lambda_{n+1}+(n-2)\mu_n\mu_{n-1}\bigg)\\\sim\Dp^2 \label{eq:err1},
\end{multline}
where the last relation is due to \cref{eq:DpControl}. This implies that as soon as $\frac1{\text{number of trajectories}}$ is comparable to $\Delta p^2$, the error does not decrease by increasing the number of trajectories, a phenomenon observed in \cref{fig:5Main,fig:qutip}. We also remark that \cref{eq:err1} shows the local error of the means. In the case of fixed timestep, the global error can grow up to $O(\delta t)$ as well.

\section{Python snippets}
\label{app:pythonSnippets}
A Python function using the Pythonic interface of C++QED to run a single trajectory may read:{\tiny\begin{verbatim}
from cpypyqed import *

def runSingleTraj(cutoff,delta,eta,kappa,nTh,n0,Dt,T,dpLimit,seed) :
    p=parameters.ParameterTable()

    pe=evolution.Pars(p)
    pm=mode.ParsPumpedLossy(p)
    
    pm.cutoff=cutoff; pm.delta=delta; pm.eta=eta; pm.kappa=kappa; pm.minitFock=n0; pm.nTh=nTh

    pe.Dt=Dt; pe.T=T; pe.dpLimit=dpLimit; pe.seed=seed

    pe.evol=evolution.Method.SINGLE; pe.dc=0; pe.epsAbs=1e-12; pe.epsRel=1e-06

    evolve(mode.init(pm),mode.make(pm,QMP.IP),pe)
\end{verbatim}}
This is equivalent to the \texttt{PumpedLossyMode\_C++QED} script mentioned in \cref{sec:code}, and indeed uses the same C++ libraries.

The much simplified algorithm in the case presented in \cref{sec:pureDpControl}, when the MCWF is reduced to a random walk in Fock space can be implemented in Python as follows:{\tiny\begin{verbatim}
import random
import numpy as np

def calculateRates(dp,nTh,n) :
    r0=2*(nTh+1)*n # photon-loss rate
    r1=2*nTh*(n+1) # photon-absorption rate
    return r0,r1,dp/(r0+r1)

def performStep(n,dtToDo,nTh,dp) :
    r0,r1,dtTryNext=calculateRates(dp,nTh,n)
    ran=random.random()/dtToDo
    if ran<r0 : nextState=n-1; jump=0 # photon loss occurs
    elif ran<r0+r1 : nextState=n+1; jump=1 # photon absorption occurs
    else : nextState=n; jump=-1 # no jump occurs
    return nextState,dtTryNext,jump
\end{verbatim}}


\bibliographystyle{unsrtnat}
\biboptions{sort&compress}

\end{document}